# Demonstration of Tunable Displacement-Measurement-Sensitivity using Variable Group Index in a Ring Resonator

*G.S. Pati, M. Salit, K. Salit, and M.S. Shahriar*
*EECS Department, Northwestern University,*
*Evanston, IL 60208*

## Abstract

We show that an intra-cavity medium with normal dispersion reduces the sensitivity of the cavity resonance frequency to a change in its length by a factor inversely proportional to the group index. Since the group index in an atomic medium can be very large, this effect can help in constructing highly frequency-stable cavities for various potential applications without taking additional measures for mechanical stability. The results also establish indirectly the opposite effect of enhanced sensitivity that can be realized for a negative dispersion corresponding to a group index close to a null value. This enhancement in turn can be employed to increase significantly the sensitivity of a ring laser gyroscope.

**PACS numbers**: 42.5, 42.50.Ct, 42.50.Gy, 42.60.Da

Optical cavities are used in many diverse applications such as optical gyroscopes, laser frequency stabilization, cavity ring-down spectroscopy, and atom-field interactions for quantum information processing[1,2,3,4]. The response of an optical cavity to the input field can be significantly modified by using a resonantly dispersive intra-cavity medium. In a recent study, we discussed a cavity-based optical gyroscope consisting of a dispersive medium, where the shift in the cavity resonance frequency due to a rotation varies inversely as the group index associated with the medium dispersion. In particular, we showed that the sensitivity of the gyroscope is enhanced by many orders of magnitude using an intra-cavity negative dispersion with the group index approaching a null value, a condition known as the critically anomalous dispersion (CAD)[5]. We have already realized experimentally a cavity with an intra-cavity medium near the CAD condition.[6] However, it was difficult to demonstrate the enhancement in sensitivity directly using this apparatus. This is primarily because the enhancement in sensitivity near the CAD condition becomes extremely large, so that the effect would be evident only for a very small change in the cavity length. To circumvent this constraint, significant improvements to the parameters of the experiment have to be made before the enhancement can be demonstrated directly near the CAD condition.

While this effort is underway, in this paper we report the results of an experiment that confirms the essential aspect of this enhanced sensitivity. Specifically, we demonstrate that the shift in the resonance frequency as a function of a change in the length of the cavity depends inversely on the group index. This experiment is performed using a positive dispersion medium instead of a CAD medium. A positive dispersion can be produced by many resonant processes involving atomic coherence induced in multi-level atomic systems. For our study, electromagnetically induced transparency (EIT)[7] or Raman gain[8] processes are used in a long path length ring cavity containing a rubidium vapor cell. The effect of reduced absorption and steep linear dispersion near the two-photon resonance on the cavity response has been studied earlier[9,10,11] by using an intra-cavity EIT medium. However, none of these studies have noted nor observed the effect of the medium dispersion on the cavity sensitivity. Our experiment shows for the first time that a positive dispersion produces a shift in the medium-cavity resonance which is highly insensitive to the cavity length change. We demonstrate quantitatively that the reduction factor in sensitivity is given by the inverse of the group index, in agreement with our theoretical model.

As mentioned above, the key purpose of this experiment is to establish indirectly the validity of the enhancement in sensitivity under the CAD condition. However, the reduced-sensitivity demonstrated here may find applications of its own. For example, the insensitivity achieved by a very large group index may be useful in forming highly stable cavities. Such cavities may find potential applications in laser frequency stabilization, producing squeezed light for gravitational wave interferometers[12], and in other applications that rely on narrow linewidth, low loss optical cavities immune to external perturbations.

In our experiment, a 10 cm long rubidium vapor cell is used inside a 100 cm long ring cavity. The cavity consists of four mirrors, with two partially transmitting plane mirrors that generate pairs of input and output ports, and two concave mirrors that allow cavity mode-matching. The empty cavity finesse is measured to be close to 100 and the cavity linewidth is measured to be about 3 MHz. Fig. 1a shows a schematic of our experimental set up. One of the concave mirrors is attached to a piezoelectric transducer (PZT) and is used in adjusting the length of the cavity. One pair of cavity input and output ports was used to lock the cavity length at a desired resonant frequency.

For the intra-cavity medium, we used a three level Λ-type system in the $D_2$ transition lines of $^{85}$Rb vapor, as shown in fig.1b. For generating intra-cavity dispersion, we have the option to choose either EIT or off-resonant Raman gain. The probe and the pump beams for the experiment are obtained from a CW Ti:Sapphire laser (line width ~ 1 MHz), with its frequency locked to a saturated-absorption atomic resonance in a separate vapor cell. The difference frequency between them is matched to the ground state splitting in $^{85}$Rb (3.0357 GHz). The probe beam is aligned to resonate in the cavity and is combined with a co-propagating, orthogonally-polarized pump using an intra-cavity polarizing beam splitter (PBS). A second intra-cavity PBS is used to separate the pump from the probe before detection. It is also used in introducing simultaneously an optical pumping beam for the gain experiment. This beam is obtained from a separate tapered amplifier diode laser and is locked to the peak of the transition between $5S_{1/2}$, F = 2 and $5P_{3/2}$, F′ = 3 (fig. 1b), using saturated absorption in a third vapor cell.

All the beams to the system are delivered using optical fibers. The magnetically shielded vapor cell is wrapped with bifilarly wound coils for heating. The probe frequency is scanned using an acousto-optic modulator (AOM) in a double-pass configuration. For both the EIT and the gain experiments, the optical power in the probe field outside the cavity was chosen to be ~ 100 μW. The pump intensity was set nearly 10 times higher than the intra-cavity probe intensity. While using the Raman gain, a strong optical pump beam is used to ensure optical pumping over a large velocity group of atoms. A flipper mirror was used in the cavity beam path to interrupt the cavity resonance, in order to monitor, when necessary, the probe field under the EIT (or the gain) condition.

The experiment is carried out as follows. First, the probe frequency is set to satisfy the two-photon resonance condition. The cavity is then made resonant at this probe frequency by adjusting its length $L$ with the PZT. The cavity length is actively held fixed at this value by using the cavity output produced by a resonating lock beam (see fig.1). The frequency of this lock beam is set at multiples of the cavity free spectral range (FSR) away from the probe frequency, $\omega_o$. The optical path of the lock beam is chosen to be different from that of the probe.

The linewidth of the cavity containing the vapor cell, in the absence of the EIT, is measured to be about 8 MHz. This is nearly three times broader than the empty cavity linewidth. This broadening is attributable primarily to the absorption by the intra-cavity medium. The effect of the dispersion under the EIT is observed by switching the pump beam on. As shown in fig. 2a, the EIT-affected cavity linewidth (~ 1.5 MHz) is narrower by a factor of nearly five than the cavity linewidth without dispersion. Note that this is also narrower than the empty cavity linewidth (3 MHz), as expected from the theoretical model[5]. Similar linewidth narrowing has also been observed in earlier experiments[7,8]. For a medium of length $\ell$, the cavity linewidth narrowing can be predicted from the dispersion-induced change of phase (or dephasing) $\delta\varphi = (\partial\varphi/\partial\omega).\delta\omega$ for a deviation $\delta\omega$ away from the resonant frequency $\omega_o$, with the phase expressed as $\varphi = \omega.(L-\ell)/c + \omega.n(\omega).\ell/c$. The expression for the modified cavity linewidth $\delta\omega'_{1/2}$ (FWHM) in the presence of the medium dispersion can be written as[11]:

$$\frac{\delta\omega'_{1/2}}{\delta\omega_{1/2}} = \left| (\sin^{-1}\left[\frac{1-R\rho}{2\sqrt{R\rho}}\right] / \sin^{-1}\left[\frac{1-R}{2\sqrt{R}}\right]) / [1 + (n_g - 1).\frac{\ell}{L}] \right|, \quad n_g \equiv 1 + \omega.\frac{\partial n}{\partial\omega}\bigg|_{\omega=\omega_o}, \quad \rho \equiv e^{-\frac{\alpha\ell}{2}} \quad (1)$$

where $\delta\omega_{1/2}$ is the empty cavity linewidth, $n_g$ is the group index, R is the reflectivity of each of the beam-splitters, and $\alpha$ is defined as the overall loss coefficient. For simplicity, the mean index of refraction for the medium at $\omega = \omega_o$ is assumed to be unity. If dispersion is negligible (i.e., $n_g$=1), then the linewidth is broadened due to the loss induced attenuation. To see the effect of dispersion, consider first the simplest case of $L=\ell$, for which the linewidth narrowing is inversely proportional to the group index $n_g$. When $\ell < L$, the behavior of the linewidth broadening is qualitatively the same, except that the value of $n_g$ needed to achieve the same degree of narrowing is larger by a factor close to $(L/\ell)$. Note that narrowing occurs for any positive dispersion: $n_g > 1$. It can also occur for negative dispersion under certain conditions. For $L=\ell$, the necessary condition is that the negative dispersion has to be steep enough so that $n_g < (-1)$. For $\ell < L$, the negative dispersion has to be steeper yet, so that $n_g < (1-2L/\ell)$.

Figure 2b shows a simulation of the cavity response illustrating the linewidth narrowing. This is obtained by substituting an analytic expression[5, 13] for the refractive index $n(\omega)\left(=\sqrt{1+\text{Re}[\chi]}\right)$ and the absorption co-efficient $\alpha(=[\mu_o\omega^2/k^2]\text{Im}[\chi])$, where $\chi$ is the medium susceptibility for the probe beam. A close match to the observed broadening is obtained for the estimated values of the experimental parameters such as the Rabi frequencies for the pump and the probe fields, the optical depth, and the cavity parameters. We also observed a similar narrowing effect when the system is operated for Raman gain on the probe (by turning on the optical pumping beam). In this case, the narrowing is a consequence of both the dispersion and the gain (since $\rho > 1$ in this case, with $\alpha$ being positive), as can be seen from eqn. 1.

Next, we describe the effect of the cavity length variation on its resonance frequency. Figure 2a shows the shift in the center of cavity resonance both in the presence and in the absence of the medium dispersion effect, as the cavity length $L$ is changed. $L$ is reduced (increased) by simply red-(blue-) detuning the lock frequency $\omega_c$ away from its original value, while the cavity is still operating under the active servo. In the absence of the medium dispersion, the center of cavity resonance gets shifted by $\Delta\omega = \omega_c - \omega_o$. Here, this frequency shift directly corresponds to a cavity length change $\Delta L$. However, in the presence of the medium dispersion, the center of cavity resonance frequency is shifted by a *smaller* amount $\Delta\omega_o'$ from $\omega_o$ (two-photon resonant frequency for the medium). Both the experimental and simulation results in fig. 2 show that this shift changes its sign by either red- or blue- detuning the lock condition which is equivalent to increasing or decreasing the cavity length from its original length $L$. The magnitude of this shift is a measure of the cavity sensitivity to its length change. This is quantified by calculating the change in the resonance frequency for a fractional change $\Delta L$ in the cavity length, using the cavity resonance condition $\omega = 2\pi N c/(n(\omega)L)$, (where N is a large integer $O(L/\lambda)$), and is given by

$$\Delta\omega_o' = \Delta\omega_o / [1 + (n_g - 1) \cdot (\ell/L)] \qquad (2)$$

where $\Delta\omega_o$ and $\Delta\omega_o'$ are the shifts in the cavity resonance frequency for the empty cavity and the loaded cavity, respectively. The frequency shift, similar to the linewidth narrowing, is inversely proportional to the group index $n_g$, and this effect is also scaled nearly by $(L/\ell)$, the ratio of the cavity length to the medium length. This effect can be considered as an indirect manifestation of the frequency-pulling effect which is described theoretically in earlier references[9,11]. However, it has neither been studied experimentally nor discussed in the context of sensitivity modification.

Figure 3 shows a sequence of such experimental results when the resonance frequency of the cavity is gradually changed from $\omega_o$ by changing the lock frequency. Two effects are observed. First, the peak amplitude of the transmitted signal for the loaded-cavity resonance is reduced with increasing $\Delta\omega_o$ (or $\Delta L$). This is because at the new resonance frequency, the probe now does not satisfy the two-photon resonance condition for EIT. In particular, if the shift $\Delta\omega_o'$ (corresponding to a given $\Delta\omega_o$) exceeds the EIT linewidth, the signal level corresponding to maximum probe transmission for the loaded-cavity resonance is considerably reduced. Second, for increasing $\Delta\omega_o$, the reduced shift $\Delta\omega_o'$ is no longer determined strictly by the expression in eqn. 2. This is because eqn. 2 assumes the variation of the index to be linear. In practice, of course, the linearity assumption fails to hold once $\Delta\omega_o'$ becomes comparable to the EIT linewidth (~ 1 MHz). A more general expression for the reduced shift $\Delta\omega_o'$ can be obtained by going beyond this assumption. It can be found by solving the following self-consistent equations for $\Delta\omega_o'$:

$$\Delta\omega_o' = \frac{\Delta\omega_o}{1 + [n_{g,eff}(\Delta\omega_o') - 1] \cdot \frac{\ell}{L}}; \quad n_{g,eff}(\Delta\omega_o') \equiv 1 + \omega_o' \left[\frac{n(\omega_o') - n(\omega_o)}{\Delta\omega_o'}\right] = 1 + (\omega_o + \Delta\omega_o') \left[\frac{n(\omega_o + \Delta\omega_o') - 1}{\Delta\omega_o'}\right] \qquad (3)$$

To understand this expression, consider first the simple case where $L = \ell$. The linewidth reduction factor is then given by the inverse of the *effective* value of the group index, which is determined by the difference of the index value at the frequency $\omega_o'$ corresponding to the shifted loaded-cavity resonance, and the index value ($\cong 1$) at the central frequency $\omega_o$. For $\ell < L$, a larger value of $n_{g,eff}$ is needed to get the same effect, just as explained in the case of eqn. 2. An exact solution of this equation can be found graphically by finding the point of intersection between the following pair of equations plotted as a function of $x(\equiv \Delta\omega_o')$:

$$y = \frac{\Delta\omega_o}{1 + (\omega_o + x)[(n(\omega_o + x) - 1)/x] \cdot (\ell/L)}; \quad y = x \tag{4}$$

From the physical interpretation of eqn. 3, it is clear that, for positive dispersion, the magnitude of $\Delta\omega_o'$ eventually gets smaller and smaller with increasing $\Delta\omega_o$, as the effective group index becomes smaller and smaller away from the line center. The linewidth at the shifted resonance is still given by eqn. 1, except that $n_g$ is replaced by the *local* group index at $\omega_o'$, defined as $n_{g,local} = 1 + \omega_o'(\partial n/\partial \omega)|_{\omega_o'}$ [5]. Figure 4 shows experimentally measured values of the reduced shift $\Delta\omega_o'$ as a function of $\Delta\omega_o$. Here, the range of $\Delta\omega_o$ is chosen such that $\Delta\omega_o'$ is restricted to within the EIT linewidth. The slope of the linear fitting shows reduced sensitivity of the cavity nearly by the same factor as the theoretically expected sensitivity reduction factor, $S$, defined as $S \equiv 1 + (n_g - 1) \cdot (\ell/L) \approx 5$. Experimentally, it should be possible to achieve a much larger value of $S$ accompanied simultaneously by linewidth narrowing, via optimization of the experimental parameters. In particular, the factor of $(\ell/L)$ can be increased to unity if one uses a cavity filled entirely by the medium. Similarly, the value of $n_g$ can be made much higher by using a buffer-gas filled cell to decrease the linewidth.

This effect may be potentially useful for several applications. For example, consider a laser with the gain medium as well as an EIT medium inside the cavity. If we assume that the bandwidth of the gain medium is much broader than that of the EIT medium, then the dispersion of the gain medium can be neglected. Under such a scenario, the laser frequency will be highly insensitive to length fluctuations caused by vibration, temperature variations, etc. A cavity of this type has also been proposed for frequency dependent squeeze amplitude attenuation and squeeze angle rotation for gravitational wave detection (GWD)[12]. In reference 12, it has been pointed out that such a cavity has potentially significant advantages over other techniques due to the narrowness and tunability of the linewidth, coupled with low optical losses. The result we are presenting here point out an additional advantage of this filter in that it is highly insensitive to fluctuations in the position of the cavity mirrors due to vibrations or temperature fluctuations.

While the experimental result presented here is for positive dispersion only, the theoretical model is valid for negative dispersion as well. We already described above the conditions under which negative dispersion also causes linewidth narrowing and reduced sensitivity. However, there is a range of negative dispersion for which the effect is reversed.

For simplicity, let us restrict the discussion to the case where $\ell = L$. In that case, if the dispersion is such that $|n_g| < 1$, the linewidth is broadened, and the sensitivity is enhanced. The effect is most pronounced at the CAD condition where $n_g = 0$. As can be seen from eqn. 1, the width of the cavity resonance becomes infinite in this case. Similarly, as can be seen from eqn. 2, the sensitivity enhancement also becomes infinite. The divergence is a result of the assumption that the index variation is linear around $\omega_o$ for all frequencies. Once the true behavior of the index variation is taken into account, it can be shown that the cavity linewidth is large but finite[5]. Similarly the sensitivity enhancement is also large but finite[5]. The enhanced bandwidth under this condition has been proposed for constructing the so-called white-light-cavity (WLC)[14,15], which can be useful for enhancing the sensitivity-bandwidth product for GWD, for example. The WLC effect has been demonstrated using a whispering gallery mode cavity[16]. However, this device is unsuitable for application to GWD[6]. We have demonstrated[6] a version of the WLC that is suited for GWD, using a variation of the experiment we have described above. Specifically, we used a dual-peaked Raman gain to realize the negative dispersion necessary for this demonstration.

The enhanced sensitivity under the CAD condition may be used to increase the signal for a given rotation rate (i.e., enhanced sensitivity) in a ring resonator gyroscope. This is because the effect of rotation can be shown to be equivalent to a change in the cavity length, as shown in ref. 5. However, for a passive resonator, this does not lead to an actual enhancement in the capability of a rotation sensor, which is characterized by the minimum measurable rotation rate. This results from the fact that while the signal level is enhanced, the linewidth broadening compensates for it, so that there is no net improvement in the sensitivity. However, as we have discussed in detail in ref. 5, this is not the case if an active resonator (i.e., a ring laser gyroscope: RLG) is used. To see why, note that for an RLG, the linewidth depends only on the cavity decay time[17], which is unaffected by the CAD condition[14,15,17]. Thus, in order to demonstrate fully that the CAD condition can be used for ultra-precise rotation sensing, it is necessary to realize an RLG with a built-in CAD medium. This is a difficult challenge, and efforts are underway in our laboratory to realize such a device.

While a passive cavity can not be used to improve the performance of a rotation sensor, it should nonetheless be possible to use it to establish the idea that the CAD condition leads to an enhancement in sensitivity. Thus, one might expect that the experimental apparatus used by us to observe the WLC effect under the CAD condition in ref. 6 could easily be used to demonstrate this enhancement. However, there are some significant difficulties in doing so, for the following reasons. The enhancement factor is non-linear: it decreases with increasing values of the empty-cavity frequency shift, $\Delta\omega_o$ (corresponding to $\Delta L$, or equivalently, a rotation rate)[5]. Furthermore, in order for the enhancement to be evident, the value of the loaded-cavity frequency shift, $\Delta\omega_o'$ (i.e., the enhanced shift) must be less than the dispersion bandwidth. Thus, for the limited dispersion bandwidth realized in ref. 6, one must use a very small value of $\Delta L$ in order to observe a significant enhancement. This in turn requires the use of a resonator that has a much higher finesse than the one used in ref. 6, and a more precise voltage supply for the PZT. The required modifications are non-trivial, and efforts are underway in our laboratory to implement these changes to the apparatus.

In conclusion, we have shown experimentally that dispersion due to an intra-cavity medium modifies the sensitivity of the cavity resonance frequency to a change in its length by a factor which is inversely proportional to the group index. Since the group index under atomic coherence can be made extremely large, the sensitivity of a long path length optical cavity can be reduced significantly. This can help in constructing highly frequency-stable cavities for various potential applications without taking additional measures for mechanical stability. The results also establishes indirectly the opposite effect of increased sensitivity that can be realized for a negative dispersion corresponding to a group index close to a null value, with potential application to ultra-precision rotation sensing. This work was supported in part by the Hewlett-Packard Co. through DARPA and the Air Force Office of Scientific Research under AFOSR contract no. FA9550-05-C-0017, and by AFOSR Grant Number FA9550-04-1-0189.

Fig.1 (a) Schematic of experimental set-up (b) two-photon transitions used in $^{85}$Rb D$_2$ line for EIT (solid lines) and Raman gain (dashed lines)

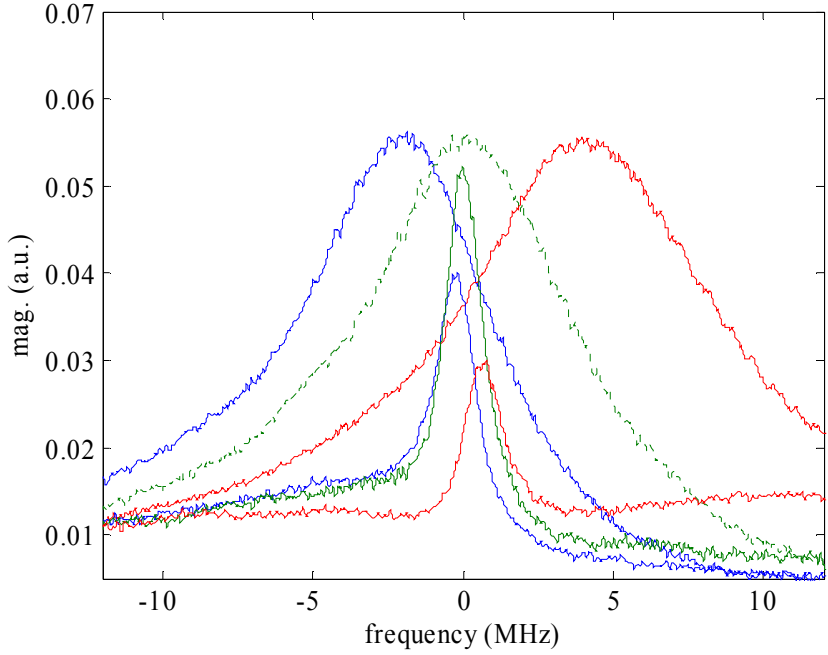

**(a)**

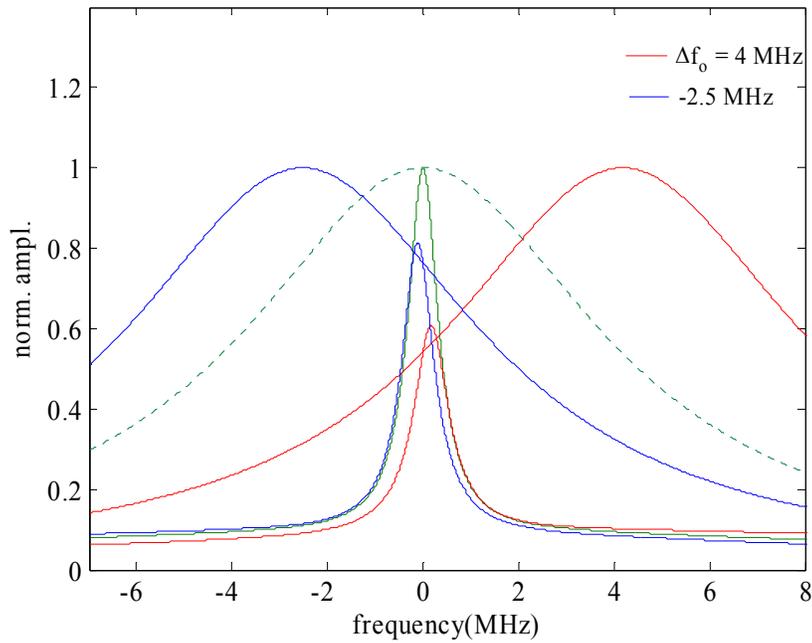

**(b)**

Fig.2 (a) Experimental results showing linewidth narrowing and reduced shift in cavity resonance for $\Delta f_o$ = -2.5, 0 and 4 MHz respectively (b) theoretical model for cavity response (Cavity linewidth = 8 MHz, EIT Linewidth = 1 MHz, $L/\ell$ = 10 and $n_g \approx 50$)

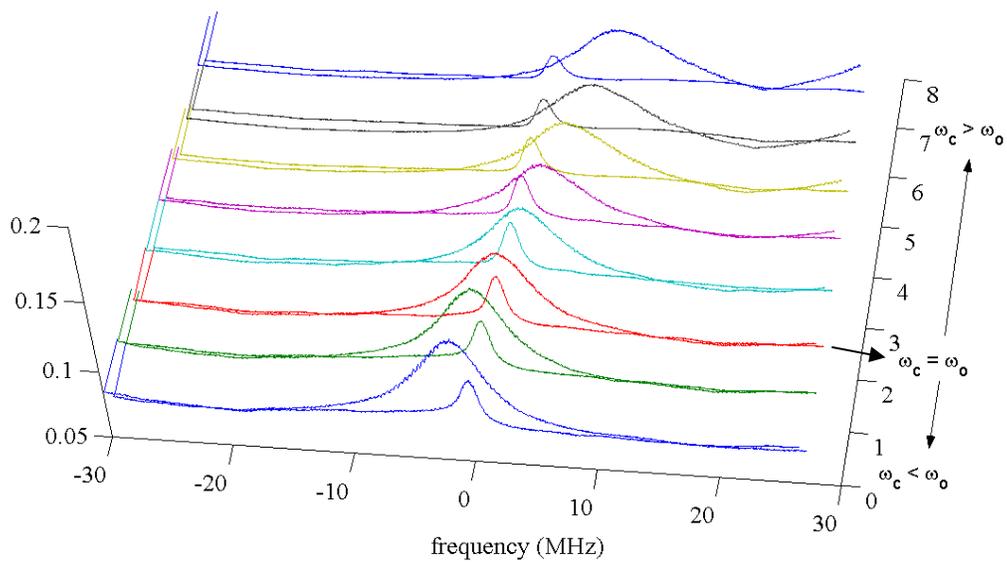

Fig.3 A sequence of cavity transmissions with (narrow resonances) and without (broad resonances) the effect of the intra-cavity medium dispersion for detuned cavities, illustrating reduced sensitivity to length changes.

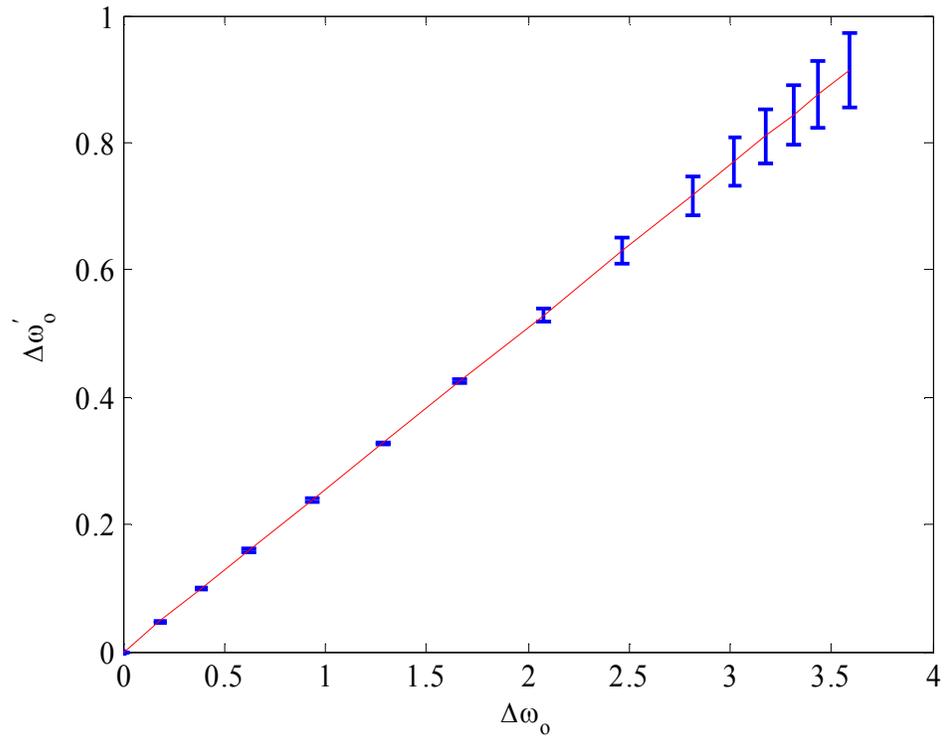

Fig.4  Experimental data illustrating how the reduced frequency shift (vertical axis) depends on the frequency shift expected without the intra-cavity medium (horizontal axis). The straight line superimposed on the data represents the theoretical model of eqn. 2, with good agreement.